\documentclass[fleqn,twoside]{article}
\usepackage[headings]{espcrc2}
\usepackage{amssymb}

\readRCS
$Id: espcrc2.tex,v 1.2 2004/02/24 11:22:11 spepping Exp $
\usepackage{graphicx}

\newcommand{\AmS}{{\protect\the\textfont2
  A\kern-.1667em\lower.5ex\hbox{M}\kern-.125emS}}

\title{Transverse Spin Effects at COMPASS}

\author{C. Schill\\[3mm] {\it on behalf of  the COMPASS collaboration}\\[3mm]
Physikalisches Institut der Albert-Ludwigs
Universit\"at Freiburg, \\Hermann-Herder Str. 3\\  
        D-79104 Freiburg, Germany. \\[2mm]   }%

\runtitle{Transverse Spin Effects at COMPASS}
\runauthor{C. Schill}

\begin{document}

\begin{abstract} The investigation of transverse spin and transverse momentum
effects in deep inelastic scattering  is one of the key physics programs of the
COMPASS collaboration. In the years 2002-2004 COMPASS took data scattering
160~$GeV$ muons on a transversely polarized $^6LiD$ target. In 2007, a
transversely polarized $NH_3$ target was used. Three different channels to
access the transversity distribution function have been analyzed: The azimuthal
distribution of single hadrons, involving the Collins fragmentation function,
the azimuthal dependence of the plane containing hadron pairs, involving the
two-hadron interference fragmentation function, and the measurement of the
transverse polarization of $\Lambda$ hyperons in the final state. Transverse
quark momentum effects in a transversely polarized nucleon have been
investigated by measuring the Sivers distribution function.  Azimuthal
asymmetries in unpolarized semi-inclusive deep-inelastic scattering give
important information on the inner structure of the nucleon as well, and can be
used to estimate both the quark transverse momentum $k_T$ in an unpolarized
nucleon and to access the so-far unmeasured Boer-Mulders function. COMPASS has 
measured these asymmetries using spin-averaged $^6LiD$ data.
\vspace{1pc} \end{abstract}

\maketitle

\section{Introduction} Most of our knowledge of the inner structure of the
nucleon is encoded in parton distribution functions. They are used to describe
hard scattering processes involving nucleons. While there has been achieved a
lot of understanding concerning the longitudinal structure of a fast moving
nucleon, very little is known about its transverse structure
\cite{Anselmino0}. Transverse refers to the direction of motion and concerns
both the transverse spin distribution and the parton intrinsic motion, $k_T$. 

Recent data on single spin asymmetries in semi-inclusive deep-inelastic
scattering (SIDIS) off transversely polarized nucleon targets
\cite{COMPASS,HERMES} triggered a lot of interest towards the transverse
momentum dependent and spin dependent distribution and fragmentation functions
\cite{Bacchetta}. Correlations between spin and transverse momentum ($k_T$)
induce new spin effects, which would be zero in the absence of intrinsic motion
of the quarks in the nucleon \cite{Aram}.  

The SIDIS cross-section in the one-photon exchange approximation contains eight
transverse-momentum dependent distribution functions \cite{Bacchetta3}. Some of
these  can be extracted in SIDIS measuring the azimuthal distribution of the
hadrons in the final state \cite{Aram2}. Three distribution functions  survive
upon integration over the transverse momenta: These are the quark momentum
distribution $q(x)$, the helicity distribution $\Delta q(x)$, and the
transversity distribution $\Delta_T q(x)$ \cite{Collins}. The latter is defined
as the difference in the number density of quarks with momentum fraction $x$
with their transverse spin parallel to the nucleon spin and their transverse
spin anti-parallel to the nucleon spin \cite{Artru}. 

To access transversity in SIDIS, one has to measure the quark polarization,
i.e. use the so-called 'quark polarimetry'. Different techniques have been
proposed so far.  Three of them are used in COMPASS: \begin{itemize} \item a
measurement of the single-spin asymmetries (SSA) in the azimuthal distribution
of the final state hadrons (the Collins asymmetry) \item a measurement of the
SSA in the azimuthal distribution of the plane containing  final state hadron
pairs (the two-hadron asymmetry) \item a measurement of the polarization of
final state hyperons (the $\Lambda$-polarimetry) \end{itemize}

Of special interest among the transverse-momentum dependent distribution
functions are  the Sivers function $\Delta_0^Tq(x, \overrightarrow{k_T})$
\cite{Sivers}, which describes a possible deformation in the distribution of
the quark intrinsic transverse momentum in a transversely polarized nucleon
\cite{Bacchetta4}, and its chiral-odd partner, the  Boer-Mulders function
\cite{Boer}, describing the transverse parton polarization in an unpolarized
hadron. 

The Boer-Mulders function generates azimuthal asymmetries in
unpolarized SIDIS, together with the so-called Cahn effect \cite{Cahn}, which
arises from the fact that the kinematics is non-collinear when $k_T$ is taken
into account, and with the perturbative gluon radiation, resulting in higher
order QCD processes.

In the past, unpolarized azimuthal asymmetries have been measured by the EMC
collaboration~\cite{emc1,emc2}, with a liquid hydrogen target and a muon beam,
without separating hadrons of different charge. These data have been
used~\cite{Aram} to extract the average $\langle k_\perp^2 \rangle$ of
quarks in the nucleon. Azimuthal asymmetries have been also measured by
E665~\cite{e665} and at higher energies by ZEUS~\cite{zeus}.   More recent are
the  measurements done by HERMES \cite{giordano} and the COMPASS results
presented here.

\section{The COMPASS experiment}

COMPASS is a fixed target experiment at the CERN SPS accelerator with a wide
physics program focused on the nucleon spin structure and on hadron
spectroscopy. COMPASS investigates transversity and the transverse momentum
structure of the nucleon in semi-inclusive deep-inelastic scattering. A
$160$~GeV muon beam is scattered off a transversely polarized $NH_3$ or
$^6LiD$  target. The scattered muon and the produced hadrons are detected in a
wide-acceptance two-stage spectrometer with excellent particle identification
capabilities \cite{Experiment}.  In the years 2002, 2003, and 2004, data were
collected  on a transversely polarized $^6LiD$ target. In the run $2007$,  
data were taken with a transversely polarized $NH_3$ target. In the following I
will focus on the new results from the data taken on the polarized $NH_3$ 
target and on the results for unpolarized azimuthal asymmetries. The latter
were obtained by averaging over opposite spin orientations of the $^6LiD$
target.

\section{The Collins asymmetry}

In semi-inclusive deep-inelastic scattering the transversity
distribution $\Delta_Tq(x)$ can be measured in combination with the
chiral odd Collins fragmentation function
$\Delta^0_TD_q^h(x)$. According to Collins, the
fragmentation of a transversely polarized quark into an unpolarized
hadron generates an azimuthal modulation of the hadron distribution 
with respect to the lepton scattering plane \cite{Collins}. The hadron
yield $N(\Phi_{Coll})$ can be written as:
\begin{equation}
N(\Phi_{Coll})=N_0\cdot (1+f\cdot P_t\cdot D_{NN}\cdot A_{Coll}\cdot \sin\Phi_{Coll}),
\label{equ:Collins}
\end{equation}
where $N_0$ is the average hadron yield, $f$ the fraction of
polarized material in the target, $P_t$ the target polarization, 
$D_{NN}=(1-y)/(1-y+y^2/2)$ the depolarization factor, and $y$ the fractional 
energy transfer of the muon. The angle $\Phi_{Coll}$ is the
Collins angle. It is defined as  $\Phi_{Coll}=\phi_h+\phi_s-\pi$, the sum of the
hadron azimuthal angle $\phi_h$ and the target spin vector azimuthal angle
$\phi_s$, both with respect to the lepton scattering plane \cite{Artru}. The
measured Collins asymmetry $A_{Coll}$ can be factorized into a convolution of the
transversity distribution $\Delta_Tq(x)$ and the
Collins fragmentation function $\Delta_T^0D_q^h(z, p_T)$, summed over all quark
flavors $q$:
\begin{equation}
A_{Coll}=\frac{\sum_q\,  e_q^2\cdot \Delta _Tq(x)\cdot \Delta_T^0D_q^h(z, p_T)}
{\sum_q\, e_q^2 \cdot q(x)\cdot D_q^h(z, p_T)}.
\end{equation}
Here, $e_q$ is the quark charge, $D^h_q(z, p_T)$ the unpolarized fragmentation
function, $z=E_h/(E_\mu-E_{\mu'})$ the fraction of available energy carried by
the hadron and $p_T$ the hadron transverse momentum with respect to the
virtual photon direction. As can be seen from equation~(\ref{equ:Collins}), the
Collins asymmetry shows as a $\sin\Phi_{Coll}$ modulation in the number of produced
hadrons.

\section{Two-hadron asymmetry}

The chiral-odd transversity distribution $\Delta_T q(x)$ can be measured
in combination with the chiral-odd polarized two-hadron interference fragmentation 
function $H^{\sphericalangle}_1 (z,M^2_{inv})$ in SIDIS. $M_{inv}$ is the invariant mass of the
$h^+h^-$ pair. 
The fragmentation of a transversely polarized quark into two unpolarized
hadrons leads to an azimuthal modulation in $\Phi_{RS} = \phi_R + \phi_s -
\pi$ in the SIDIS cross section. 
Here $\phi_R$ is the azimuthal angle between $\vec R_T$ and the lepton scattering plane and 
$\vec R_T$ is the transverse component of $\vec R$ defined as:
\begin{equation}
\vec R = (z_2\cdot \vec p_1 - z_1 \cdot \vec p_2)/(z_1+z_2).
\end{equation}
 $\vec p_1$ and $\vec p_2$ are the momenta in the laboratory frame of $h^+$
and $h^-$ respectively. This definition of $\vec R_T$ is invariant
under boosts along the virtual photon direction.

The number of produced oppositely charged hadron pairs $N_{h^+h^-}$ can be written as:
\begin{equation}
N_{h^+h^-} =N_0 \cdot ( 1 + f \cdot P_t \cdot D_{NN} \cdot A_{RS} \cdot \sin \Phi_{RS} \cdot \sin
\theta).
\end{equation}
Here, $\theta$ is the angle between the momentum vector of $h^+$ in
the center of mass frame of the $h^+h^-$-pair and the momentum vector of
the two hadron system \cite{Bacchetta}. 

\begin{figure}
     \includegraphics[width=0.9\columnwidth]{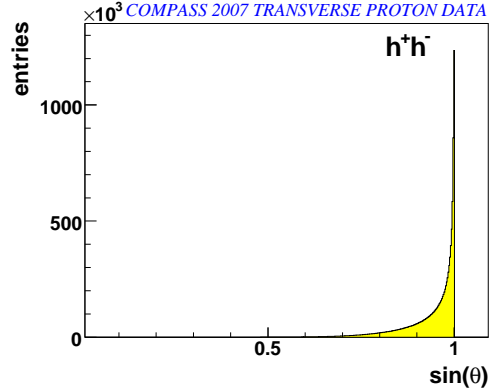}
     \vspace*{-1.2cm}
     \caption{$\sin \theta$ distribution}
     \label{pic:sin_theta}
\end{figure}

The measured amplitude $A_{RS}$ is proportional to the product of the
transversity distribution and the polarized two-hadron interference fragmentation function 
\begin{equation}
A_{RS} \propto \frac {\sum_q e_q^2 \cdot \Delta_T q(x) \cdot H^{\sphericalangle}_1(z,M^2_{inv})}
 {\sum_q e_q^2 \cdot q(x) \cdot D_q^{2h}(z,M^2_{inv})}.
\end{equation}
$D_q^{2h}(z,M^2_{inv})$ is the unpolarized two-hadron interference fragmentation function.
The polarized two-hadron interference fragmentation function can be expanded in the relative
partial waves of the hadron pair system, which up to the
p-wave level gives~\cite{Bacchetta}:
\begin{equation}
H^{\sphericalangle}_1 = H^{\sphericalangle,sp}_1 + \cos \theta H^{\sphericalangle,pp}_1.
\end{equation}
Where $H^{\sphericalangle,sp}_1$ is given by
the interference of $s$ and $p$ waves, whereas the function
$H^{\sphericalangle,pp}_1$ originates from the interference of two $p$
waves with different polarization. For this analysis the results are
obtained by integrating over $\theta$, because the $\sin \theta$
distribution shown in Fig.~\ref{pic:sin_theta} is strongly peaked at one
and the $\cos \theta$ distribution is symmetric around zero. 

Both the interference fragmentation function
$H_1^\sphericalangle(z,M_{inv}^2)$ and the corresponding spin averaged fragmentation
function into two hadrons $D_q^{2h}(z, M_{inv}^2)$ are unknown, and need to be measured in $e^+e^-$
annihilation or to be evaluated using models \cite{Bacchetta,Jaffe,Bianconi,Radici}. 

\section{The Sivers asymmetry} 

Another source of azimuthal asymmetry is related to the Sivers effect. The
Sivers asymmetry rises from a coupling of the intrinsic transverse
momentum $\overrightarrow{k}_T$ of unpolarized quarks with the spin of a
transversely polarized nucleon \cite{Sivers}. The correlation between the transverse nucleon
spin and the transverse quark momentum is described by the Sivers distribution
function  $\Delta_0^Tq(x, \overrightarrow{k}_T)$. The Sivers effect leads to an
azimuthal modulation of the produced hadron yield:
\begin{equation}
N(\Phi_{Siv})=N_0\cdot (1+f\cdot P_t\cdot A_{Siv}\cdot \sin \Phi_{Siv}).
\label{equ:Sivers}
\end{equation}
The Sivers angle is defined as $\Phi_{Siv}=\phi_h-\phi_s$. The measured Sivers
asymmetry $A_{Siv}$ can be factorized into a product of the Sivers distribution function and the
unpolarized fragmentation function $D_q^h(z)$:
\begin{equation}
A_{Siv}=\frac{\sum_q\, e_q^2\cdot \Delta_0^Tq(x, \overrightarrow{k}_T)\cdot  D_q^h(z)}
{\sum_q \,e_q^2\cdot  q(x)\cdot D_q^h(z)}.
\end{equation}
In this case the asymmetry $A_{Siv}$ shows up as the amplitude of a $\sin\Phi_{Siv}$ modulation in the
number of produced hadrons. 

Since the Collins and Sivers asymmetries are independent azimuthal modulations
of the cross section for semi-inclusive deep-inelastic scattering \cite{Boer}, both
asymmetries can be determined experimentally from the same dataset. 

\section{Data sample and event selection} The polarized $NH_3$ target consists of three
cells (upstream, central and downstream) of 30, 60 and 30 cm length,
respectively. The upstream and downstream cell are polarized in one direction
while the middle cell is polarized oppositely. The target material has a high
polarization of about $90$\%. The dilution factor of the ammonia target is constant at $0.15$ in
$z$ and $p_T$ bins, while it increases with $x$ from $0.14$ to $0.17$. The direction of
the target polarization was reversed every five days. The asymmetries are
analyzed using at the same time data from two time periods with opposite
polarization and  from the different target cells. The data have been selected
requiring a good stability of the  spectrometer within one and between
consecutive periods of data taking. 

To select DIS events, kinematic cuts of the squared four
momentum transfer $Q^2>1$~(GeV/c)$^2$, the hadronic invariant mass
$W>5$~GeV/c$^2$ and the fractional energy transfer of the muon $0.1<y<0.9$ were
applied. The hadron sample on which the single hadron asymmetries are computed consists of all
charged hadrons originating from the reaction vertex with $p_T> 0.1$~GeV/c
and $z>0.2$.

The Collins and Sivers asymmetries were evaluated as a function of $x$, $z$, and
$p_T$ integrating over the other two variables.  The
extraction of the amplitudes is then performed fitting the expression for the
transverse polarization dependent part of the semi-inclusive DIS cross section \cite{Boer}
to the measured count rates in the target cells by a unbinned extended maximum likelihood fit,
taking into account the spectrometer acceptance. The results have been checked
by several other methods described in Ref.~\cite{COMPASS}.

The hadron pair sample consists of all oppositely charged
hadron pair combinations originating from the reaction vertex. The
hadrons used in the analysis have $z > 0.1$ and $x_F > 0.1$. Both cuts
ensure that the hadron is not produced in the target
fragmentation. To reject exclusively produced $\rho^0$-mesons, a cut on
the sum of the energy fractions of both hadrons was applied
$z_1+z_2<0.9$. Finally, in order to have a good definition of the
azimuthal angle $\phi_R$ a cut on $R_T > 0.07$\,GeV/c was applied.

\section{Transverse target results}

\begin{figure*}
\vspace*{-0.5cm}
\centerline{
\includegraphics[width=0.95\textwidth]{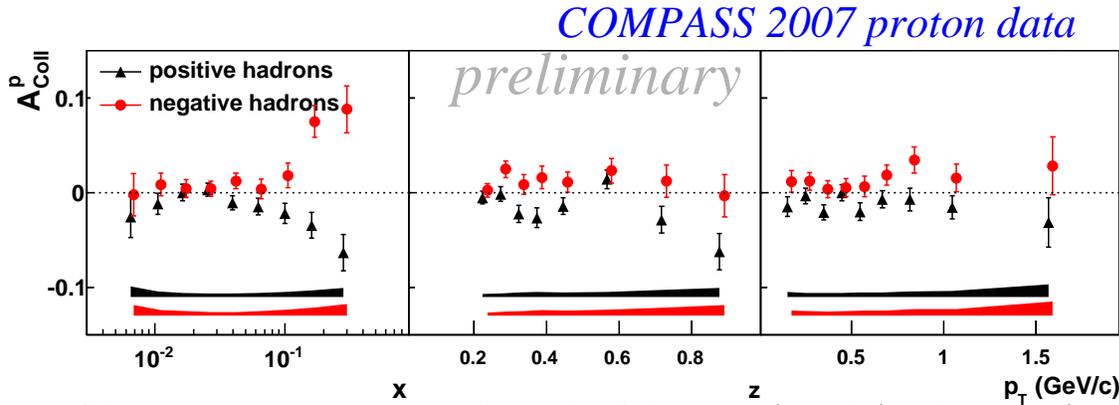}}
\vspace*{-1.5cm}
\caption{Collins asymmetry on the proton for unidentified positive (triangles) and negative
(circles) hadrons as a function of $x$, $z$, and $p_T$. The bands indicate the
systematic uncertainty of the measurement.}
\label{Collins}
\end{figure*}

\begin{figure*}
\vspace*{-0.5cm}
\centerline{
\includegraphics[width=0.95\textwidth]{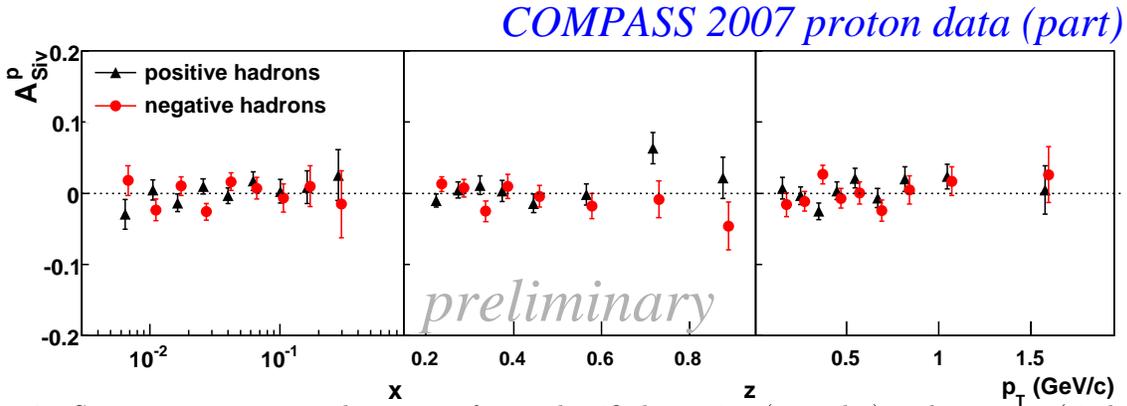}}
\vspace*{-1.5cm}
\caption{Sivers asymmetry on the proton for unidentified positive (triangles) and negative
(circles) hadrons as a function of $x$, $z$, and $p_T$.}
\label{Sivers}
\end{figure*}

\begin{figure*}
\vspace*{-0.5cm}
     \includegraphics[width=0.98\textwidth]
	{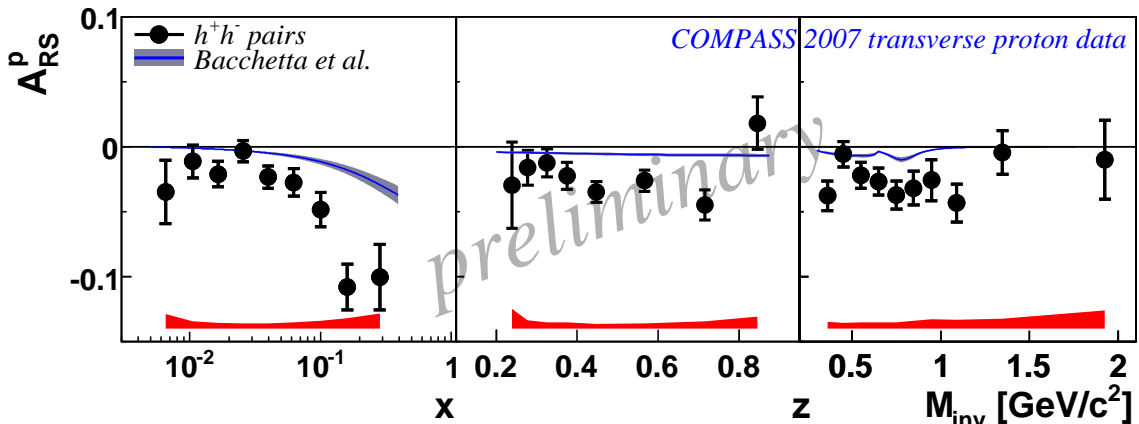}
  \vspace{-1.2cm}   
     
     \caption{Two-hadron asymmetry $A_{RS}$ as a function of $x$, $z$ and $M_{inv}$, compared to 
     predictions of \cite{Bacchetta:2008wb}. 
     The lower bands indicate the
     systematic uncertainty of the measurement.}
	\label{pic:results}

\end{figure*}

In Fig.\ref{Collins} the results for the Collins asymmetry on a proton target are
shown as a function of $x$, $z$, and $p_T$ for positive and negative  hadrons. For
small $x$ up to $x=0.05$ the measured asymmetry is small and statistically compatible
with zero, while in the last points an asymmetry different from zero is visible. The
asymmetry increases up to about $10$\% with opposite sign for negative and positive
hadrons.

In Fig.\ref{Sivers} the results for the Sivers asymmetry are shown as a
function of $x$, $z$, and $p_T$.  The Sivers asymmetry is small and
statistically compatible with zero for both positive and negative hadrons.

The two-hadron asymmetry as a function of $x$, $z$ and $M_{inv}$ is shown in
Fig.~\ref{pic:results}. A strong asymmetry in the valence $x$-region can be observed,
which implies a non-zero transversity distribution and a non-zero polarized two
hadron interference fragmentation function  $H^{\sphericalangle}_1$. In the invariant
mass binning one observes a strong signal around the $\rho^0$-mass and the asymmetry
is negative over the whole mass range. The lines are predictions from Bacchetta and
Radici, which are based on the transversity distribution from Anselmino {\it et al.}
and on a fit to HERMES data~\cite{Bacchetta:2008wb}. It can be seen that the
predictions are by about a factor of  3 smaller than the measured asymmetry.

\section{Azimuthal asymmetries in DIS off an unpolarized target}

The cross-section for hadron production in lepton-nucleon DIS $\ell N
\rightarrow \ell' h X$ for unpolarized targets and an unpolarized or
longitudinally polarized beam has the following form~\cite{bacchetta2}:

\begin{equation}
\begin{array}{lcr}
\frac{d\sigma}{dx dy dz d\phi_h dp^2_{h,T}} =
  \frac{\alpha^2}{xyQ^2}
\frac{1+(1-y)^2}{2} \cdot\\[2ex] [ F_{UU,T} + 
  F_{UU,L} + \varepsilon_1 \cos \phi_h F^{\cos \phi_h}_{UU} \\[2ex]
   + \varepsilon_2 \cos(2\phi_h) F^{\cos\; 2\phi_h}_{UU}
   + \lambda_\mu
  \varepsilon_3
  \sin \phi_h F^{\sin \phi_h}_{LU} ]
\end{array}
\end{equation}
where $\alpha$ is the fine structure constant. 
$F_{UU,T}$,  $F_{UU,L}$, $F^{\cos \phi_h}_{UU}$,  $F^{\cos\;
  2\phi_h}_{UU}$ and $F^{\sin \phi_h}_{LU}$ are structure functions. Their 
first and second subscripts indicate the beam and target polarization,
respectively, and the last subscript denotes, if present, the
polarization of the virtual photon.  $\lambda_\mu$ is the 
longitudinal beam polarization and: 
\begin{equation}
\begin{array}{rcl}
\varepsilon_1 & = & \frac{2(2-y)\sqrt{1-y}}{1+(1-y)^2} \\[2ex]
\varepsilon_2 & = & \frac{2(1-y)}{1+(1-y)^2} \\[2ex]
\varepsilon_3 & = & \frac{2 y \sqrt{1-y}}{1+(1-y)^2}
\end{array}
\end{equation}
are depolarization factors. The Boer-Mulders parton distribution function
contributes to 
the $\cos \phi_h$ and the $\cos 2\phi_h$ moments as well, together with the
 Cahn effect~\cite{Cahn} which arises from the fact that the kinematics is
non collinear when
the $k_\perp$ is taken into account, and with the
perturbative gluon radiation, resulting in order $\alpha_s$ QCD processes. pQCD
effects become important for high transverse momenta $p_T$ of the produced
hadrons.

\section{Analysis of unpolarized asymmetries}

The event selection requires standard DIS cuts, i.e. $Q^2 > 1 $ (GeV/c)$^2$,
mass of the final hadronic state $W>5 $ GeV/c$^2$, $0.1 < y < 0.9$, and the
detection of at least one hadron in the final state. For the
detected hadrons it is also required that:

\begin{itemize}
\item the fraction of the virtual photon energy carried by the hadron is
$0.2 < z < 0.85$ to select hadrons from the current fragmentation
  region. 
\item
the hadron transverse momentum is $0.1 < p_T < 1.5 $ GeV/c  for a better determination of the
azimuthal angle $\phi_h$. 
\end{itemize}

Data taken with a longitudinally polarized and a transversely polarized target have
both been spin-averaged to obtain an unpolarized data sample.  The statistics
corresponds after all cuts to  $5\times 10^6$ positive hadrons and $4\times 10^6$
negative hadrons.

In the measurement of unpolarized asymmetries a Monte Carlo simulation is used to correct for
acceptance effects of the detector. The SIDIS event generation is performed by 
Lepto\cite{lepto}, the experimental setup and the particle interactions in the 
detectors are simulated by  COMGEANT.

The experimental acceptance as a function of the azimuthal angle $A(\phi_h)$ is then calculated as the
ratio of  reconstructed over generated events for each bin of $x$, $z$ and $p_T$ in which the
asymmetries are measured. The measured distribution, corrected for acceptance, is fitted with the
following functional form:
\[
\begin{array}{lll}
N(\phi_h) &=&N_0 \left( 1 + A^D_{\cos \phi}  \cos \phi_h + \right.  \\&& \left. A^D_{\cos 2\phi}
\cos 2\phi_h
 +  A^D_{\sin \phi} \sin \phi_h \right) 
\end{array}
\]
The contribution of the acceptance corrections to the systematic error was 
studied with care.
As the
asymmetries were extracted from data taken both with longitudinal and
transverse target configurations,  comparing the two results gives the effect of the acceptance
changes due to the different configuration of the target
magnet and to the different direction of the incoming beam (for the
transverse setup the beam is bent in order to leave the target with the same direction
as in the longitudinal case). In order to check the effect of the Monte Carlo simulation
parameters, the acceptance was calculated using two different sets
of Lepto parameters. All the
resulting asymmetries were compared in order to quantify the
systematic error in each kinematic bin. Further systematic tests,
like splitting the data sample according to the event
topology and to the time of the measurement, gave no significant contributions.

\section{Results for unpolarized asymmetries}

\begin{figure}
\begin{center}
\includegraphics[width=\columnwidth]{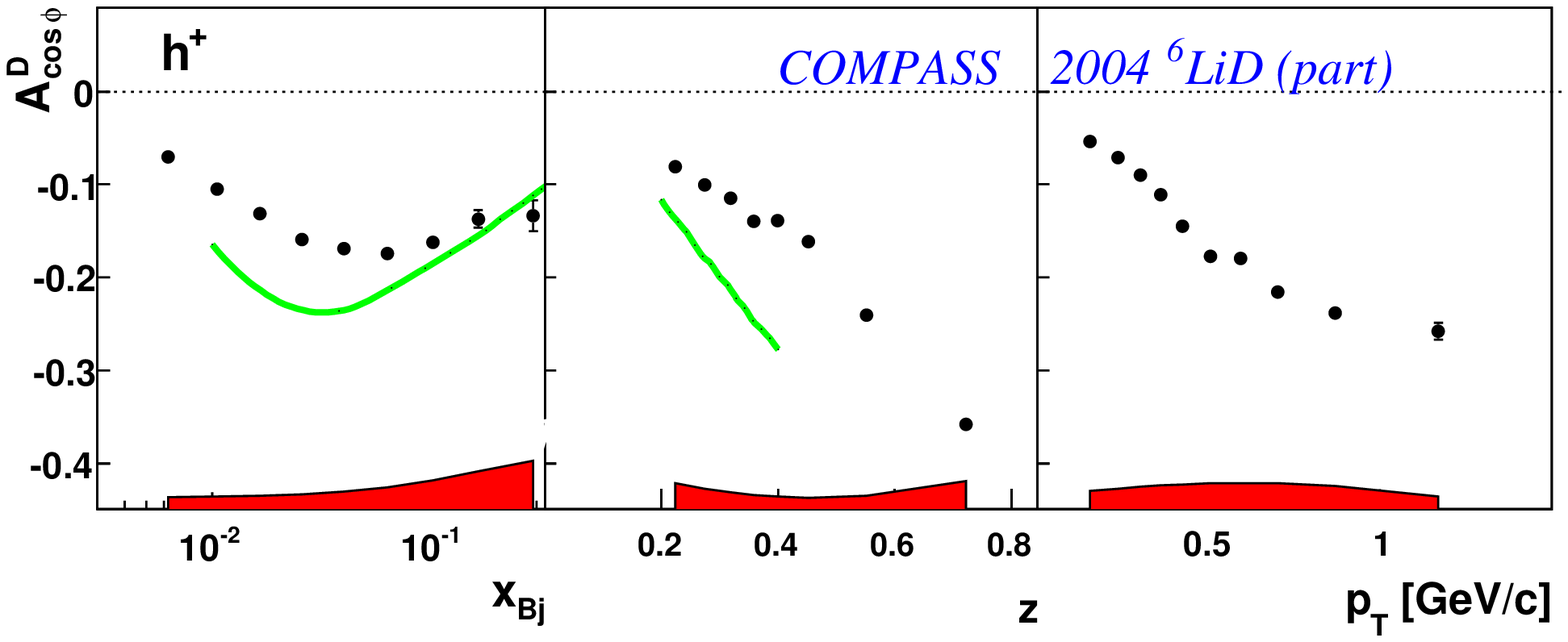}
\includegraphics[width=\columnwidth]{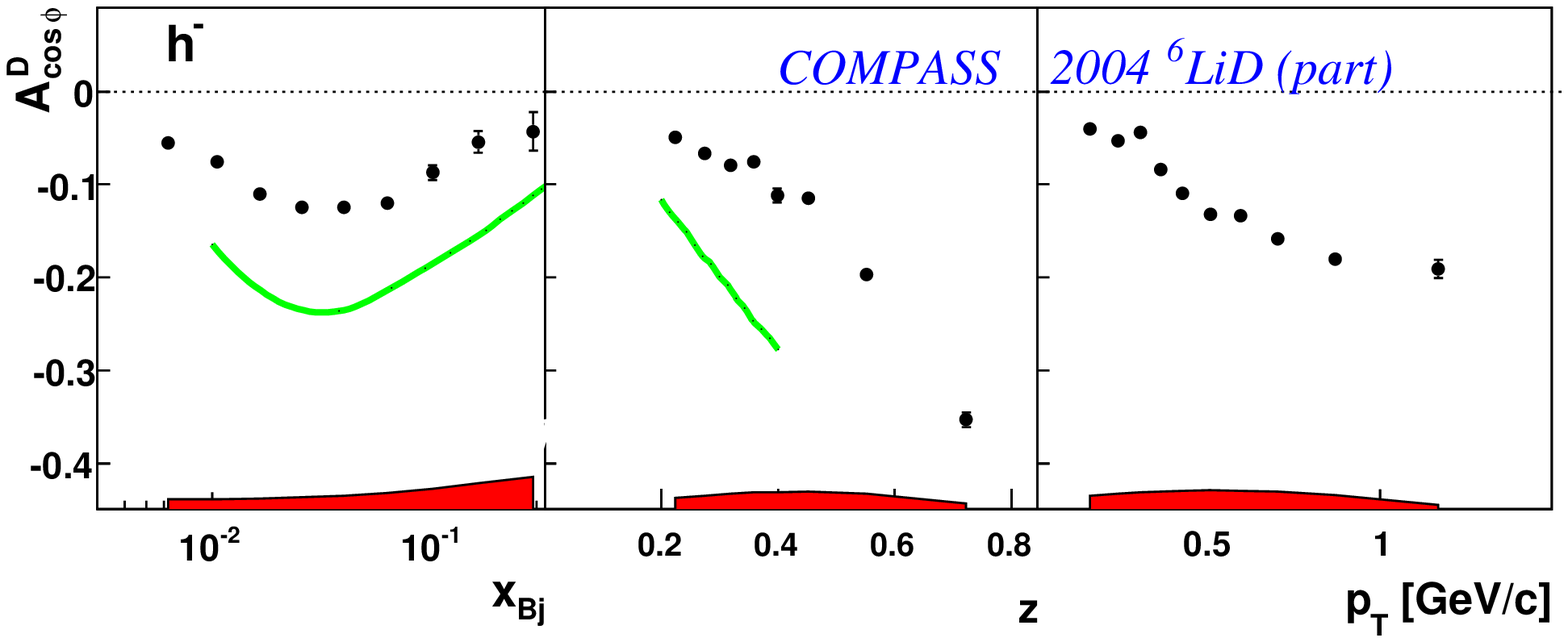}
\end{center}
\vspace{-1cm}
\caption{$\cos \phi_h$ asymmetries from COMPASS deuteron data
for positive (upper row) and negative (lower
raw) hadrons; the asymmetries are divided by the kinematic factor
$\varepsilon_1$ and the bands indicate the size of the systematic uncertainty. 
The superimposed curves are the values predicted by~\protect\cite{anselmino2}
taking into account the Cahn effect only.
}
\label{f:cosphi}
\end{figure}

The $\sin \phi_h$ asymmetries measured by COMPASS, not shown here,  are compatible
with zero, at the present level of statistical and systematic errors, over the
full range of $x$, $z$ and $p_T$ covered by the data.

The $\cos \phi_h$ asymmetries extracted from COMPASS deuteron data
are shown in Fig.~\ref{f:cosphi} for positive (upper row) and negative (lower
row) hadrons, as a function of $x$, $z$ and $p_T$. The bands indicate the size
of the systematic error. The asymmetries show the same trend for positive and
negative hadrons with  slightly larger absolute values for  positive hadrons. 
Values as large as 30$-$40\% are reached in the last point of the $z$ range. 
The theoretical
predictions~\cite{anselmino2} in Fig.~\ref{f:cosphi} takes into account the Cahn
effect only, which
does not depend on the hadron charge. The Boer-Mulders parton distribution function 
is not
taken into account in this prediction. 

\begin{figure}
\vspace{-20pt}
\begin{center}
\includegraphics[width=\columnwidth]{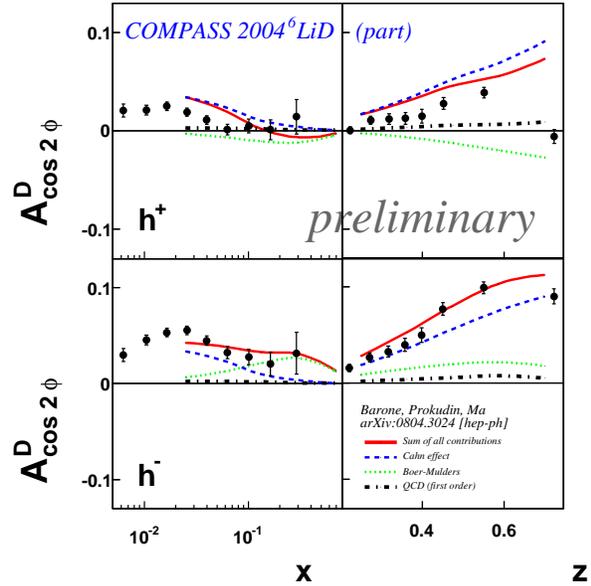}
\end{center}
\vspace{-20pt}
\caption{$\cos 2 \phi_h$ asymmetries from COMPASS deuteron data
for positive (upper row) and negative (lower
raw) hadrons; the asymmetries are divided by the kinematic factor
$\varepsilon_1$ and the bands indicate size of the systematic error. 
}
\label{f:cos2phi}
\end{figure}

The $\cos 2 \phi_h$ asymmetries are shown in Fig.~\ref{f:cos2phi} together with the
theoretical predictions of~\cite{barone}, which take into account the kinematic
contribution given by the Cahn effect, first order pQCD (which, as expected, is
negligible in the low $p_T$ region), and  the Boer-Mulders parton distribution
function (coupled to the Collins fragmentation function), which gives a different
contribution to positive and negative  hadrons. In~\cite{barone} the Boer-Mulders
parton distribution function is assumed to be proportional to the Sivers function as
extracted from preliminary HERMES data. The COMPASS data show an  amplitude
different  for positive and negative hadrons, a trend which confirms the theoretical
predictions. There is a satisfactory agreement between the data points and the model
calculations, which hints to a non-zero Boer-Mulders parton distribution function.

\section{Summary} 

New preliminary results for Collins and Sivers asymmetries measured at COMPASS in
semi-inclusive deep-inelastic scattering off a transversely polarized proton target
have been presented. For $x>0.05$, a Collins asymmetry different from zero and with
opposite sign for positive and negative hadrons has been observed. Within statistical
precision of the measurement, the Sivers asymmetry is yet compatible with zero, both
for negative and positive hadrons. In 2010, COMPASS will continue a full year of data
taking with a transversely polarized proton target and largely increase the
statistical precision of both  the Collins and Sivers results.

The measured unpolarized azimuthal asymmetries on the deuteron target show large $\cos\phi_h$ and $\cos
2\phi_h$
moments which can be qualitatively described in model calculations taking into account the Cahn
effect and the intrinsic $k_T$ of the quarks in the nucleon and the
Boer-Mulders structure function.

\end{document}